\documentclass [12pt,twoside]{article}           

\usepackage{epsfig,times,lscape}
\usepackage[usenames]{color}

    \ifnum\count102=1

    \fi

    \ifnum\count102=2
\fi

\newcommand{\nc}{\newcommand}

     \ifnum\count101=1
\nc{\qI}[1]{\section{{#1}}}
\nc{\qA}[1]{\subsection{{#1}}}
\nc{\qun}[1]{\subsubsection{{#1}}}
\nc{\qa}[1]{\paragraph{{#1}}}

\def\qpar{\vskip 2mm plus 0.2mm minus 0.2mm}
\def\qL{\hfill \break}
     \fi 

      \ifnum\count101=2
 \nc{\qI}[1]{\parindent=0mm \vskip 8mm 
{\centerline{\LARGE \color{red}#1}}\vskip 3mm}
\nc{\qA}[1]{\vskip 2.5mm \noindent {{\bf        #1}} \vskip 1mm
\parindent=0mm}
 \nc{\qun}[1]{\vskip 1mm \noindent {\sl #1 }\quad }

\def\qL{\hfill \break}
\def\qpar{\vskip 2mm plus 0.2mm minus 0.2mm}

      \fi

\def\qth{\vrule height 12pt depth 0pt width 0pt}
\def\qtb{\vrule height 0pt depth 5pt width 0pt}

\nc{\qfoot}[1]{\footnote{{#1}}}

      \ifnum\count101=1
\def\qbu{\hfill \par \hskip 6mm $ \bullet $ \hskip 2mm}
\def\qee#1{\hfill \par \hskip 6mm (#1) \hskip 2 mm}
      \fi
      \ifnum\count101=2
\def\qbu{\hfill \par \hskip 4mm $ \bullet $ \hskip 2mm}
\def\qee#1{\hfill \par \hskip 4mm (#1) \hskip 2 mm}
      \fi

\def\qparr{ \vskip 1.0mm plus 0.2mm minus 0.2mm \hangindent=10mm
\hangafter=1}

     \ifnum\count101=1 
 \def\qdec#1{\parindent=0mm\par {\leftskip=2cm {#1} \par}}
     \fi
     \ifnum\count101=2
  \def\qdec#1{\parindent=0mm \par {\leftskip=1cm {#1} \par}}
  
  \def\qcitb#1{\noindent \hbox to 102mm{\hfill \small #1} \vskip 1mm}
      \fi

 \def\qpages#1{\count102=0{\loop\advance\count102 by 1
 \null \vfill\eject \ifnum\count102<#1 \repeat}}

\def\qn#1{\eqno \hbox{(#1)}}

\def\qth{\vrule height 12pt depth 0pt width 0pt}
\def\qtb{\vrule height 0pt depth 5pt width 0pt}

\def\qv{\vskip 0.1mm plus 0.05mm minus 0.05mm}

\def\qhw{\hskip 1.5mm}
\def\qleg#1#2#3{\noindent {\bf \small #1\qhw}{\small #2\qhw}{\it \small #3}\qv }
\def\qvec#1{\overrightarrow{#1}}

\begin{document}
\thispagestyle{empty}



\markboth{{\sl \hfill  \hfill \protect\phantom{3}}}
        {{\protect\phantom{3}\sl \hfill  \hfill}}

\color{yellow} 
\hrule height 40mm depth 6mm width 170mm 
\color{black}
\vskip -4.0cm 
\centerline{\bf \LARGE \color{blue} How can one detect}
\vskip 5mm
\centerline{\bf \LARGE \color{blue} the rotation of the Earth
``around the Moon''? }
\vskip 5mm
\centerline{\bf \LARGE \color{blue} Part 1: With a Foucault pendulum}
\vskip 2cm
\centerline{\bf \large 
Bertrand M. Roehner$ ^{1,2} $ }

\large
\vskip 1cm

{\bf Abstract}\quad 
It will be shown that the rotation of
the Earth in the Earth-Moon system can be detected
by comparing the deflection of a Foucault pendulum 
at noon on the one hand and at midnight on the other hand.
More precisely, on 21 June the midnight
experiment would give a deflection about 4\% larger than at noon.
In other words, with a Foucault pendulum having an accuracy of
the order of 1\% one should be able to identify this effect 
through a single measurement. 
Moreover, if the experiment is repeated
on $ N $ successive days, the reduction of the error bar 
by a factor $ 1/\sqrt{N} $
which comes with the averaging process will allow identification
of the Moon effect even with a pendulum of poorer accuracy, say of
the order of a few percent.
In spite of the fact that this effect appears fairly easy to detect,
it does not seem that its observation
has attracted much attention so far. We hope that this paper
will encourage some new observations.
 
\vskip 8mm
\centerline{6 December 2011}
\vskip 2mm
\centerline{\it Preliminary version, comments are welcome}

\vskip 10mm
{\normalsize Key-words: rotation, Moon-Earth, Coriolis force,
detection, Foucault pendulum, gyrometer}
\vskip 20mm

{\normalsize 
1: Department of Systems Science, Beijing Normal University, Beijing,
China. \qL
2: Email address: roehner@lpthe.jussieu.fr.
On leave of absence from the ``Institute for Theoretical and
High Energy Physics'' of University Pierre and Marie Curie, Paris, France.
}

\vfill \eject

\qI{Introduction}

\qA{The Foucault pendulum seen as a gyrometer}

Improving the accuracy of experiments and 
measurement devices has been a
major means of progress in physics. 
\qpar
A device for detecting
movements of rotation is called a gyrometer or gyrosensor
or angular accelerometer.
The Foucault pendulum is a highly sensitive gyrometer%
\qfoot{There are many commercial models of gyrometers. 
However,
the sensitivity of most of them is too low for the 
kind of measurements considered here. A possible exception
may be the gyrometers used for the inertial navigation of
nuclear submarines, but (for obvious reasons) it is
almost impossible to get detailed information about
the performances of such devices.}%
.
In this paper we examine how its accuracy can be further improved.
\qpar

As is well known, the Foucault pendulum
was introduced in 1851 by the physicist
L\'eon Foucault. At that time, in spite of a length of 67 meters%
\qfoot{The rationale for using a long pendulum is related to the
so-called Puiseux effect which can be stated as follows. 
A spherical pendulum whose trajectory
is an ellipse of major axis $ 2a $ and minor axis $ 2b $ undergoes
a rotation of its major axis with an angular velocity $ \Omega_P $
given by: $ \Omega_P= (3/8)(ab/L^{5/2})\sqrt{g} $ where $ L $ is
the length of the pendulum and $ g $ the acceleration of gravity.
The Puiseux effect has exactly the same appearance as
the Foucault effect, namely a rotation of the major axis.
Therefore, if
$ ab/L^{5/2} $ is too large it
will completely hide and obliterate the Foucault effect.}%
,
the accuracy that was achieved was not better than a few percent.\qL
In his PhD thesis of 1879
the Dutch physicist Heike Kamerlingh-Onnes showed that it was
possible to build a short 
and nevertheless fairly accurate Foucault
pendulum. His pendulum had a length of about 1.50m and an accuracy
of about 1\%. In recent years as many as 18 short Foucault
pendulums based on different techniques were designed and built by
Marcel B\'etrisey%
\qfoot{The following website provides additional information
http://www.betrisey.ch/eindex.htm. It gives
useful descriptions of several of the pendulums and 
of some of the problems raised by their design and construction.
Moreover, broader information about pendulums can be
found in a recent book by Leslie Pook (2011)}%
. 
Surprisingly, however, it seems that the physical
applications of the Foucault pendulum have been restricted to
showing the movement of the Earth around its axis. This 
was not really something new even in 1851 
\qfoot{In this connection let us recall that when Galileo (allegedly)
said ``E pur si muove!'' (``And yet it moves'') he was speaking 
of the rotation of the Earth around the Sun for which
by the way,  he had no
scientific justification whatsoever. His belief
was chiefly based on the observation of Jupiter and
its satellites that he saw as a small scale model of the solar system.}
.

\qA{For what observation can the Foucault pendulum be used?}

The question raised in this paper is whether one can 
improve the accuracy of 
the Foucault pendulum and then use it to make meaningful
physical observations. The Earth is subject to many movements
of rotation, some of which are summarized in Table 1.
\begin{table}[htb]

\centerline{\bf Table 1\quad Movements of rotation of the Earth}
{\small
\vskip 3mm
\hrule
\vskip 0.5mm
\hrule
\vskip 2mm
$$ \matrix{
& \hbox{Rotation} \hfill & \hbox{Period} & \hbox{Angular} & \hbox{Precision} \cr
 &\hbox{} \hfill & \hbox{} & \hbox{velocity}  & \hbox{required} \cr
\qtb
&\hbox{} \hfill & \hbox{[day]} & \hbox{[degree/24 h]} & [\%] \cr
\noalign{\hrule}
\qth
1&\hbox{Rotation of the Earth on its axis} \hfill & 1 & 360 & \cr
2&\hbox{Rotation of the Earth ``around the Moon''} \hfill & 27 & 13 & 1 \cr
3&\hbox{Rotation of the Earth around the Sun} \hfill & 365 & 0.98 & 0.20 \cr
4&\hbox{Precession of the equinoxes} \hfill &  & 1.4\ 10^{-3} & 4\ 10^{-4} \cr
\qtb
5&\hbox{Rotation of the solar system around the center of the galaxy} \hfill &
 73\ 10^{9}& 1.3\ 10^{-11} & 4\ 10^{-12} \cr
\noalign{\hrule}
} $$
\vskip 1.5mm
Notes: The angular velocity column gives the angular deviation in 24 hours 
of the plane of
a Foucault pendulum located at the North pole as observed by a 
terrestrial observer. The last column gives the precision with which one
must measure the deviation in order to be able to detect the rotation 
mentioned in the same line of the table.
The accuracy of standard Foucault pendulum experiments is comprised
between 0.5\% and 1\% and is therefore too low
for the effects number 3, 4 and 5 to be observable.
The so-called precession of the equinoxes is a slow change in the direction
of the axis of rotation of the Earth, an effect which is similar to the phenomenon by which the axis of a spinning 
top ``wobbles'' when a torque is applied to it. \qL
In addition of the movements listed in the table our galaxy is also moving 
toward the Andromeda galaxy 
with a relative velocity equal to 220 km/s (which is about 7 times more
than the speed of the Earth on its orbit around the Sun). However, 
at the time
of writing, it is not clear whether
 this movement is also a rotation around some
(still unknown) center.
{\it Source: Roehner (2007, chapter 5)}
\vskip 2mm
\hrule
\vskip 0.5mm
\hrule
\vskip 3mm
}

\end{table}

Can we use the Foucault pendulum as a ``rotation observatory''
in order to detect some of these movements?
\qpar

After the rotation of the Earth around its axis the first possible
candidate is the movement with respect to the Moon.
We use to say that the Moon rotates around the Earth and that the
Earth revolves around the Sun, but in fact in both cases the rotation
is around the center of gravity of the two-body systems.
In the case of the
Earth-Moon system the center of gravity is in fact inside
the Earth as shown in Fig. 1.

%
\begin{figure}[htb]
 \centerline{\psfig{width=12cm,figure=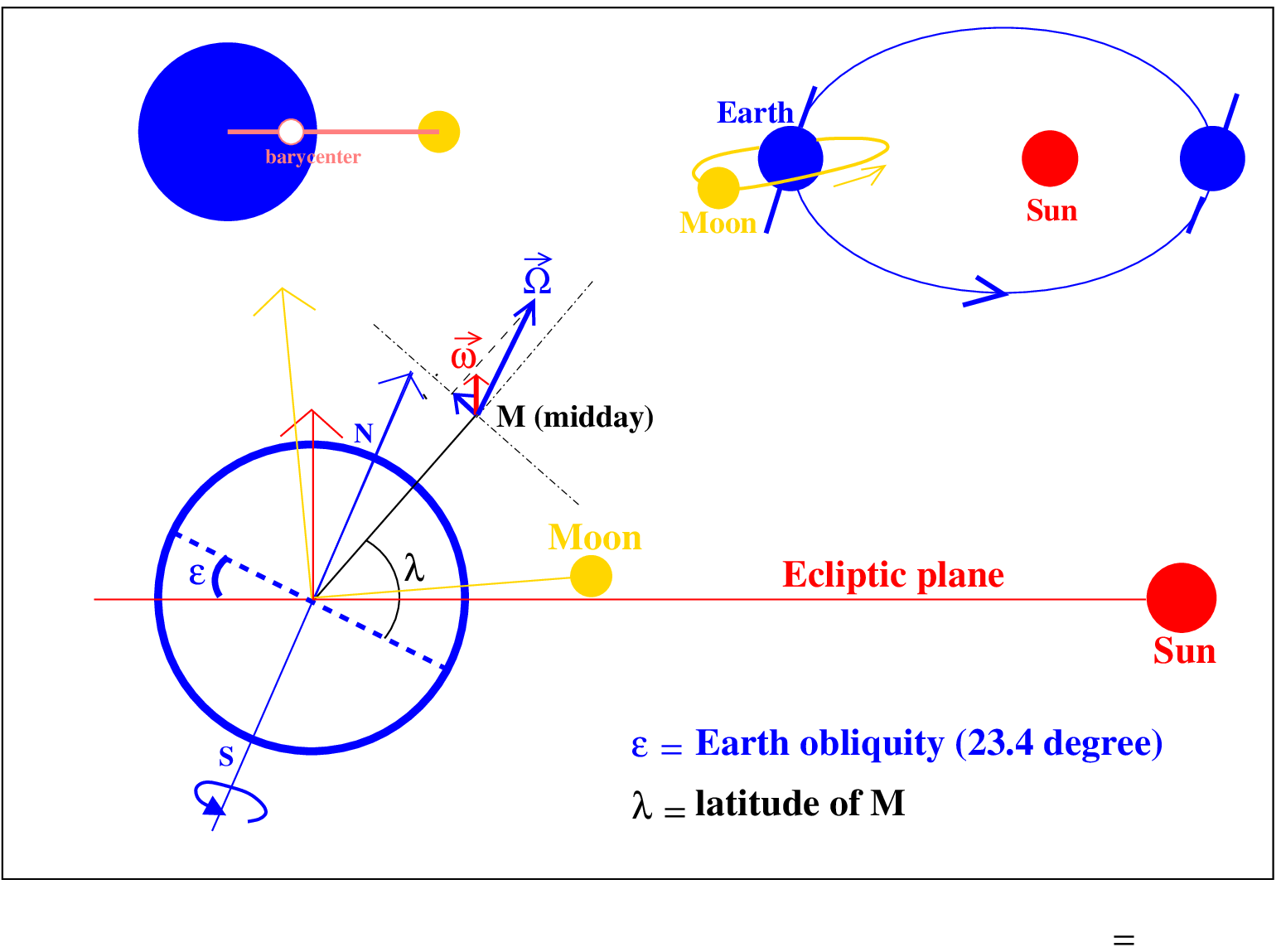}}
\vskip 3mm
\qleg{Fig. 1: Rotation of the Earth ``around''  the Moon.}
{In fact, the Earth-Moon system rotates around the center of
gravity as represented on the top-left diagram. The orbit
of the Moon makes an angle of 5 degree with the ecliptic plane
wich contains the orbit of the Earth around the Sun, as shown
in the top-right diagram. The respective vectors of angular velocity
are represented in the diagram at the bottom of the figure.
As it is the comparison between
the midnight and midday situations which is important these
two cases are shown in more detail in Fig. 2b and 2c.}
{}
\end{figure}
%

\qI{Two interpretations of the Foucault effect}

There are two ways to consider the movement of a Foucault pendulum
depending on which frame of reference one uses.
\qbu In the first approach one uses a 
geocentric frame of reference $ GC $ whose axis are stationary with respect to distant stars. At the North Pole
a Foucault pendulum is expected 
\qfoot{In fact, we do not know if the Foucault
pendulum experiment has ever been made right at the North Pole,
but this conclusion can be extrapolated from numerous observations
made at various latitudes.}
to experience an angular deflection of 360 degrees
in 24 hours%
\qfoot{In many textbooks one reads that this deflection
corresponds to a {\it sideral day}  of 23.944 hours, rather than to the
normal 24-hour day. However the difference between the two is so
small (namely 0.27\%) that the distinction is beyond observation.}%
.

In the $ GC $ frame of reference this observation can be interpreted
by saying that the plane of oscillation of the pendulum is
motionless with respect to the ``stars'' and that the Earth rotates
below it. 
\qbu The second approach is to work in a frame of reference
located on the Earth and rotating with it. Therefore, in order
to be able to use Newton's law one must add 
to the external forces the so-called Coriolis
force $ \qvec{F_c}=2m\qvec{v}\wedge\qvec{\Omega} $, where
$ \qvec{\Omega} $ is the vecteur of angular 
velocity which is parallel to the axis of rotation of the Earth, 
and
$ \qvec{v} $ is the velocity of the pendulum's mass.
At the North pole (Fig. 2a) $ \qvec{\Omega} $ is vertical
and upward; therefore any horizontal movement 
will de deflected toward the right. As a result 
the trajectory of the pendulum will be a kind of marguerite
and the plane of oscillation will turn clock-wise. 
This conclusion is in agreement with the first perspective 
because the rotation is counter-clockwise when watched from the
pendulum. \qL
At another location the vector $ \qvec{\Omega} $ will
have both a vertical and a horizontal component (Fig. 2b).
As the movement of the pendulum is almost horizontal
the horizontal component will generate a vertical force which
will be absorbed into the tension of the wire and which plays
no role therefore. It is the vertical component 
$ \omega_v=\Omega \sin\lambda $, where $ \lambda $ is the
latitude of the location, which will deflect the pendulum.
As $ \omega_v $ is smaller than $ \Omega $ the period of the
Foucault pendulum will be longer than 24 hours and given by:
$ T=2\pi/\omega_v = 24/\sin\lambda $. On the Equator
$ \omega_v $ will be zero (Fig. 2c) and there will therefore
be no Foucault effect if one restricts oneself to the rotation
of the Earth.

\qI{Detection of the movement of the Earth around the Moon}

In this section, we also take into account the two other
rotations of the Earth (numnbered as 2 and 3 in Table 1).
Basically, we need to find the vertical projection, $ \omega_v $
of the total angular velocity: 
$$ \qvec{\omega} =\qvec{\Omega} + \qvec{\omega_1} + \qvec{\omega_2} $$

where $ \qvec{\omega_1} $ and $ \qvec{\omega_2} $ are
the rotation vectors
with respect to the Moon and Sun respectively.

Let us again begin by the simple case of the North Pole (Fig. 2a).
%
\begin{figure}[htb]
\centerline{\psfig{width=8cm,figure=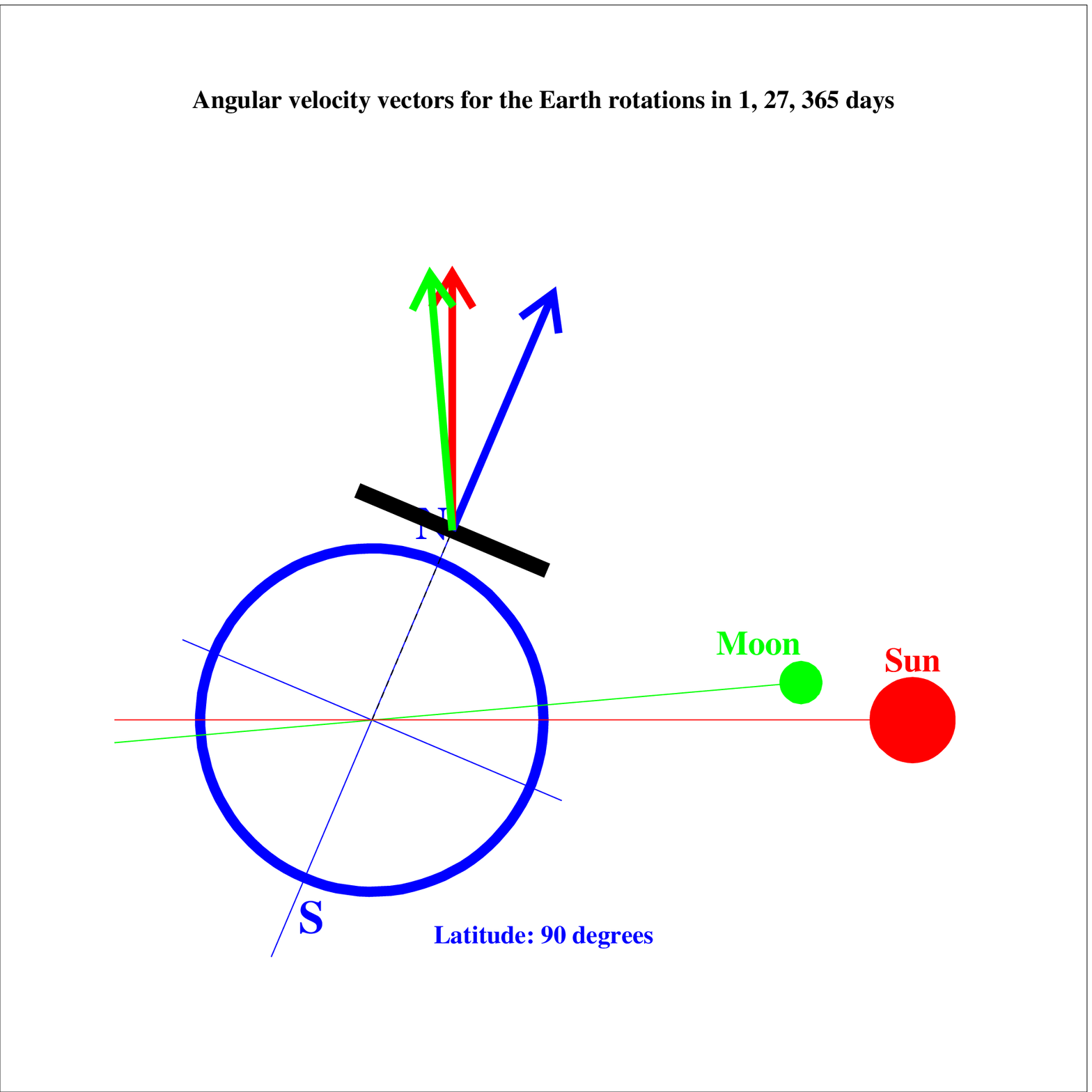}}
\vskip 3mm
\qleg{Fig. 2a: Vectors of angular velocity at the North Pole 
on 21 June (at noon).}
{For the clarity of the diagram these vectors are represented
with same length. In fact, the green vector (Moon) should be 27
times shorter than the blue vector (Earth around its axis) and the red
vector (Sun) should be 365 times shorter. 
Similarly, for the sake of clarity the figure corresponds to
the situation on 21 June. On this day the plane containing
the center of the Sun and the axis of rotation of the Earth is
perpendicular to the orbit of the Earth. For
21 June the diagram is easier
to draw but the Foucault effect is the same on other days.
For instance,
6 months later, that is to say on 21 December,
the only difference is that the North Pole
is in the night during the whole day.\qL
As the movement of the Foucault pendulum is almost horizontal,
only the vertical components of the vectors play a role.
Without the green and red rotation-vectors, the
period of the Foucault pendulum would be one day.
These additional rotations will accelerate the
Foucault deflection and make its period shorter.
However, at the North Pole (in contrast to other latitudes)
there will be no change in the Foucault effect during 
the course of one day.
As will be seen in Fig. 2c, the most drastic change of the Foucault
effect in the course of one day occurs on the Equator.}
{}
\end{figure}
%

%
\qbu In this situation the angle between $ \qvec{\omega_2} $ and
$ \qvec{\Omega} $ is equal to the angle 
between the Equator of the Earth
and the ecliptic, that is to say $ \epsilon=24.5 $ degrees. 
We denote by $ m $ the angle between the  orbit of the Moon and
the ecliptic: $ m=5 $ degrees. Then the angle between the vertical
and $ \qvec{\omega_1} $ is $ \epsilon + m $. 
In this case there is no change of the total vertical
projection of $ \qvec{\omega} $ in the course of one day.
\qbu At another location of latitude $ \lambda $ the 
vertical projection of 
$ \qvec{\omega}_1  + \qvec{\omega}_2 $ will change in the
course of one day (Fig. 2b). 
%
\begin{figure}[htb]
 \centerline{\psfig{width=8cm,figure=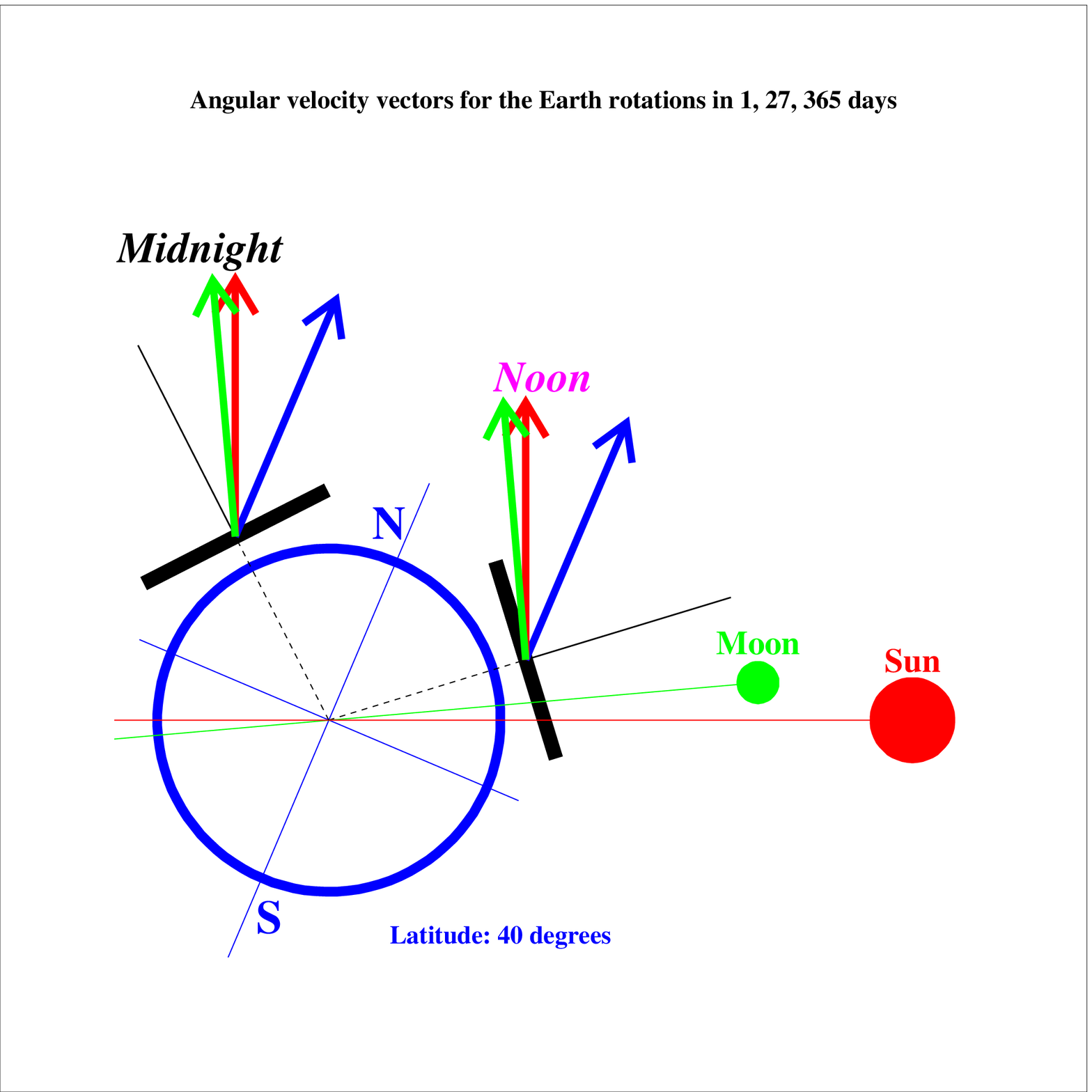}}
\vskip 3mm
\qleg{Fig. 2b: Comparison of the Coriolis force at noon and
midnight at a latitude of $ \lambda = 40 $ degrees (on 21 June).}
{As (for a given mass and velocity)
the Coriolis force $ \qvec{F_c}=2m\qvec{v}\wedge\qvec{\omega} $
is basically determined by the vector of
angular velocity $ \qvec{\omega} $,
one needs to compare these vectors at noon and midnight.
The vertical projection of the blue vector is the same
at noon and midnight, namely $ \Omega \sin\lambda $. 
On the contrary, the vertical projections of the green and red vectors,
will be different. The diagram shows that the projection is greater
at midnight than at noon. By using
the two expressions given in the text 
the difference is seen to be about 4.4\%.
On 21 December it is the opposite: at noon the Foucault
effect is larger than at midnight by 4.4\%.}
{}
\end{figure}
%
Indeed, whereas the vertical projection of $ \qvec{\Omega} $
will always be $ \Omega \sin{\lambda} $, the vertical
projection of  $ \qvec{\omega}_1 $ and  $ \qvec{\omega}_2 $
will change as the Earth turns around its axis.
In Fig. 2b which corresponds to 21 June, the two extreme cases
occur at noon and midnight. At another date, the diagram would
be similar, only the hours are not the same. Thus, on 21 March
these two extreme cases occur six hours later.\qL
In the situation shown in Fig. 2b which corresponds to a
latitude of 40 degrees, the total vertical components 
at noon and midnight will have
the following expressions%
\qfoot{For a different latitude these expressions 
are slighly modified depending on the specific position of
the vertical with respect to the vectors $ \qvec{\omega_1} $ and
$ \qvec{\omega_2} $.}%
:
$$ \omega_v^{(12)}=\Omega \sin\lambda
+ { \left( \Omega \over 27 \right) }\sin(\lambda -m-\epsilon)
+ { \left( \Omega \over 365 \right) }\sin(\lambda -\epsilon) \qn{1a} $$

$$ \omega_v^{(24)}=\Omega \sin\lambda
+ { \left( \Omega \over 27 \right) }\sin(\lambda+m+\epsilon)
+ { \left( \Omega \over 365 \right) }\sin(\lambda +\epsilon) \qn{1b} $$

For $ \lambda = 40 $ degrees, these formulas lead to:
$$ { \omega_v^{(24)}-\omega_v^{(12)} \over \omega_v^{(12)} }=4.4\% $$

In other words, there will be a difference of 4.4\% between 
measurements carried out at midnight and noon respectively.
\qbu The difference between noon and midnight is much more
drastic at the Equator (Fig. 2c)
%
\begin{figure}[htb]
 \centerline{\psfig{width=8cm,figure=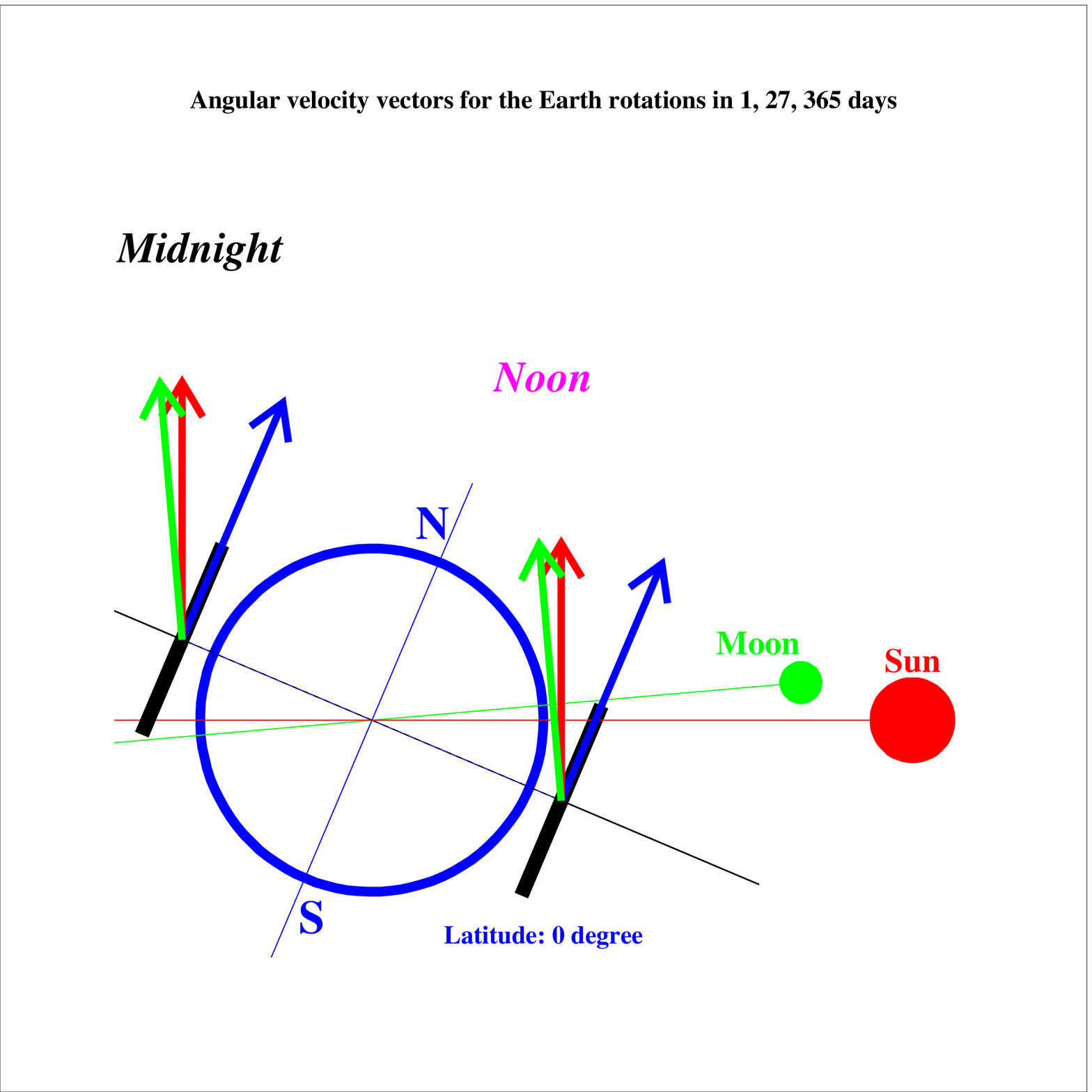}}
\vskip 3mm
\qleg{Fig. 2c: Comparison of the Coriolis force at noon and
midnight on the Equator (on 21 June).}
{This case is interesting because the behavior of the Foucault
pendulum is drastically different at noon and
midnight. This comes from the fact that on the Equator the
effect due to the rotation of the Earth is 
inexistant (the vertical component of the blue vector is zero).
Therefore, the Foucault effect is determined solely by the projections
of the green and red vectors. The fact that these projections
are of opposite sign (upward at midnight and downward at noon)
means that the corresponding
deflections of the Foucault pendulum are in opposite 
directions. \qL
On 21 December 
the diagram is the same with the only difference that noon and
midnight are interchanged.}
{}
\end{figure}
%
because in this case $ \qvec{\Omega} $
does not contribute to the vertical projection. The
figure shows that the vertical projections of $ \qvec{\omega_1} $
and $ \qvec{\omega_2} $ is upward at midnight and downward at noon
which means that the Foucault pendulum will be deflected
clockwise at midnight and counter-clockwise at noon. 
Of course, the angular velocity is nuch slower.
A whole rotation will take $ 27/\sin(23.5+5)= 56 $ days
\qpar

In the next section we examine how the previous predictions can
be tested experimentally.

\qI{How to set up experimental tests}

It seems that in recent decades most of the Foucault pendulums
which have been constructed were more destined to decoration
purposes
than to making physical measurements. 
Because they must be able to 
move permanently these pendulums
included an electromagnetic device
for maintaining the oscillations. 
Of course, this device introduces
external forces which may slighly change the behavior of the
pendulum. This leads us to suggest that 
for physical measurements such devices should be 
disconnected. As the tests described below require only fairly short
measurements over time intervals of a few minutes, 
such devices will not be necessary anyway%
\qfoot{It may be argued that a ``passive'' (that is to say
without any source of current) Charron ring should not change 
the behavior of the pendulum. Indeed, its purpose is to prevent the
trajectory from becoming more elliptic, thus keeping the 
behavior of the pendulum closer to the theoretical model.
This can be beneficial but on the other hand may
not be necessary for within a 
few minutes the ``ovalisation''
will remain limited. So, one should probably try the two
options, with and without Charron ring, in order to see what
difference it makes.}%
.

\qA{Measurement of the angular deflection}
At the latitude of 40 degrees, the angular deflection 
of an ideal Foucault pendulum will be
0.16 degree per minute. In order to make such an angle measurable
with reasonable accuracy, it must be amplified. 
Several methods can be proposed. A simple method is to take pictures
which will then be magnified on a computer screen.\qL
One can also design an optical method.
Such a method should work even when the mass of the pendulum 
rotates around the direction of the wire because this is what
will indeed occur except for pendulums which have a rigid suspension. 
Moreover, the measurement should not be affected by the damping
of the oscillations. The following method satisfies these two
conditions. 
\qee{1} One uses a cylindrical mass in order to be able to 
fix a circular mirror onto the upper surface of the cylinder.
\qee{2} In the position where
 the pendulum is at the end of its trajectory a
well-focused laser beam is directed toward the mirror.
This beam comes from above with an angle of about 45 degree.
\qee{3} After the beam has been reflected upward a reflexion on
a second mirror makes the direction of the beam almost horizontal
so that it can be sent toward a screen located at 
a distance $ D $ of the order of 10 to 20 meters.
\qee{4} What is the amplification power of such a detection
system? By conducting static trials we found that on the 
screen the deviation of the laser spot is about 5mm per
degree of rotation of the plane of oscillation of the pendulum
and per meter of distance between the second mirror and the
screen. 
\qpar

As an illustration, we consider an experiment which
lasts 2mn and produces a deflection of $ 2\times 0.16=0.32 $ degree.
With $ D=10 $m the horizontal deviation of the spot on the
screen will be about $ 5\times 0.32\times 10 = 15 $mm.\qL
When, due to damping, the amplitude of the pendulum is reduced, 
the angle of the 
circular mirror with respect to the horizontal will decrease.
As a result the spot on the scren will become lower but its
horizontal position should not be affected.

\qA{Improving the accuracy through averaging}

In physics averaging is a standard procedure for reducing the
error-bar of measurements%
\qfoot{At the time of writing (December 2011)
this method is used at the CERN Laboratory in Geneva
for deciding whether the
Higgs particle exists or not. The method simply requires
that one accumulates a sufficiently large number of events so that
the error bar can be reduced accordingly.
To record such a large number of events the LHC accelerator
must run for months.}%
. 
This procedure is based
on the fact that the standard deviation of an average of
$ N $ uncorrelated measurements $ X_k $
each of which has a standard deviation
$ \sigma $ is $ \sigma/\sqrt{N} $. 
$$ X_m= { X_1+X_2+\ldots + X_N \over N },\quad 
\sigma(X_m)={ \sigma \over \sqrt{N} } $$

Usually, one must distinguish between systematic and random errors.
Random errors may for instance be due to vibrations or changes
in friction forces.
Clearly, the reduction by $ \sqrt{N} $ applies only to random
errors. 
\qpar

However, for the comparison of noon and midnight deflections
one does not have to care about systematic errors {\it provided
they are identical at noon and at midnight.} 
On the contrary,
if the measurement is modified by a change in
temperature and if (as is likely)
the midnight temperature is systematically lower than the noon
temperature, then the comparison will be affected. The same
can be said for vibrations for they are also likely to be lower
at midnight than at noon.
\qpar

Another question which must be addressed is the following.
\qdec{Is it better to make 10 separate measurements over 2-minute
time intervals or rather one measurement over a 20-minute time 
interval?}
\qpar

To get a better insight into this question let us
consider a simpler example.\qL
When one measures the period of a pendulum with a chronometer
it is more accurate to count 20 oscillations rather
than to make 10 measurements of only 2 oscillations. Why?\qL
In this case the main source of error is the uncertainty $ \Delta $
when starting and stopping the chronometer. While the measurement
of 20 oscillations comprise an uncertainty of $ 2\Delta $, 
the 10 measurements
of two oscillations comprise an uncertainty of $ 20\Delta $; 
due to the averaging
process, this uncertainty will be reduced to $ (20/\sqrt{10})\Delta $,
but this is still larger (by a factor $ \sqrt{10} $)
than the $ 2\Delta $ uncertainty.\qL
In other words, the answer to the previous question depends on
the level of uncertainty in launching the pendulum
(at the end of the 2-minute run there is no special uncertainty).
Clearly the answer will also depend upon the magnitude and
effects of the ``ovalisation'' of the trajectory.
\qpar

\qA{Prospects}

It is difficult to know in advance which level of accuracy can
be achieved. The measurements can be repeated 25 times, 100 times
or (once the procedure has been automatized) 1,000 times.
As there is in principle no limit to the number of
repetitions, thanks to $ 1/\sqrt{N} $ factor of the averaging
process,
the accuracy of the experiment can become fairly high.
With a Foucault pendulum which has an intrisic accuracy of
1\%, $ N=1,000 $ would lead to an accuracy of 
$ 1/\sqrt{1000} = 0.03\% $. Such an accuracy would be sufficient
to observe the effect of $ \qvec{\omega_2} $, that is to say
the rotation of the Earth around the Sun. 
\qpar

However, the best chance to identify the effect of the rotation of
the Earth around the Sun is by a dual measurement done
at noon and midnight on 21 June and 21 December respectively.
The noon and midnight results should
be interverted (see above Fig. 2c). For this measurement one needs
only the 1\% accuracy that is required for observing the
rotation around the Moon.

\qI{Conclusion}

In the second of this series of two papers we will introduce a 
different kind of gyrometer. However, before
introducing a new method, it is natural to try
those who already exist. 
As the Foucault pendulum
has been studied (and possibly improved) for over 
one and a half century, it is an obvious candidate.
We have described how its accuracy can be improved,
firstly through amplification of the angular deflection,
secondly through an appropriate number of repetitions.
We have also shown what kind of observations must be done
in order to identify specific movements of rotation of the Earth.
\qpar

It appears that the accuracy required for detecting
the movement in the Earth-Moon system can be achieved fairly
easily. By automatising the measurement process 
one would be able to carry out a great
number of repetitions and in this way one may get an accuracy sufficient
for detecting the rotation of the Earth around the Sun. 
That was the dream of Galileo some four centuries ago and 
the realization of this dream would
be an appropriate tribute to the founding father of modern physics.

\vskip 10mm

{\bf References}

\qparr
B\'etrisey (M.) 2011: Foucault pendulums.\qL
http://www.betrisey.ch/eindex.htm. 

\qparr
Kammerlingh-Onnes (H.) 1879: Nieuwe bewijzen voor de aswenteling der aarde" [New proofs of the rotation of the earth].
PhD dissertation, Groningen.

\qparr
Pook (L.P.) 2011: Understanding pendulums. A brief introduction.
Springer, Berlin.

\qparr
Roehner (B.M.) 2007: Driving forces in physical, biological and socio-economic phenomena:
a network science investigation of social bonds and interactions.
Cambridge University Press, Cambridge.

\end{document}